\documentclass[aps,showpacs,twocolumn,showkeys,amsmath,amssymb,floatfix,nofootinbib,preprintnumbers]
{revtex4}
\usepackage{bm}

\begin{document}

\title{$\chi^{\vphantom\dagger}_{c0}(3915)$ As the Lightest $c\bar c s
  \bar s$ State}

\author{Richard F. Lebed}
\email{richard.lebed@asu.edu}
\affiliation{Department of Physics, Arizona State University, Tempe,
Arizona 85287-1504, USA}

\author{Antonio D. Polosa}
\email{antoniodavide.polosa@uniroma1.it}
\affiliation{Dipartimento di Fisica and INFN, Sapienza Universit\`{a}
  di Roma, P. Aldo Moro 2, I-00185 Roma, Italy}
\affiliation{CERN-TH, CH-1211 Geneva 23, Switzerland}

\date{February, 2016}

\begin{abstract}
  The state $\chi^{\vphantom\dagger}_{c0}(3915)$ has recently been
  demoted by the Particle Data Group from its previous status as the
  conventional $c\bar c$ $2 {}^3P_0$ state, largely due to the absence
  of expected $D\bar D$ decays.  We propose that
  $\chi^{\vphantom\dagger}_{c0}(3915)$ is actually the lightest $c\bar
  c s \bar s$ state, and calculate the spectrum of such states using
  the diquark model, identifying many of the observed charmoniumlike
  states that lack open-charm decay modes as $c\bar c s \bar s$.
  Among other results, we argue that $Y(4140)$ is a $J^{PC} = 1^{++}$
  $c\bar c s \bar s$ state that has been not been seen in two-photon
  fusion largely as a consequence of the Landau-Yang theorem.
\end{abstract}


\maketitle


\section{Introduction} \label{sec:Intro}

The past 13 years have been a time of remarkable growth in
experimental reports of hadronic states, particularly in the
charmonium and bottomonium sectors.  Starting with Belle's observation
of the $X(3872)$ in 2003~\cite{Choi:2003ue}, almost 30 new states with
masses lying in these regions have been reported.  Until last year's
observation of the baryonic $P_c^+$ states by
LHCb~\cite{Aaij:2015tga}, all of the observed states were mesonic.
Since this counting does not include new conventional quarkonium
states discovered in the interim, such as the $c\bar c$
$\chi_{c2}(2P)$~\cite{Uehara:2005qd,Aubert:2010ab}, all of these
states are considered {\it exotic}.  These additional exotic states
have been suggested in numerous papers to be gluon hybrids,
kinematical threshold effects, di-meson molecules, compact charmonium
embedded in a light-quark cloud ({\it hadrocharmonium}), and
diquark-antidiquark states (Ref.~\cite{Chen:2016qju} gives an
exhaustive recent review of work in these areas).

Evidence has steadily mounted that at least some of the mesonic
exotics are {\it tetraquarks}, and the baryonic exotics are {\it
pentaquarks}.  For example, the $Z^+(4430)$ state first observed in
2008~\cite{Choi:2007wga} is charmoniumlike but also charged, so that
its minimum valence quark content is $c\bar c u \bar d$.  But the
confirmation experiment by LHCb~\cite{Aaij:2014jqa} also measured a
rapid phase variation of the $Z^+(4430)$ production amplitude near the
peak mass, which is characteristic of true resonant scattering
behavior.  Similar observations were carried out for the $P_c^+$
states~\cite{Aaij:2015tga}.\footnote{The rapid phase variation alone
is not universally accepted as decisive evidence of a resonance, and
more discriminating tests have been
proposed~\cite{Pakhlov:2014qva,Guo:2015umn,Szczepaniak:2015eza,
Hanhart:2015zyp}.}

The definitive separation of exotic from conventional states is not
always trivial, however.  The $X(3872)$ has the same $J^{PC} = 1^{++}$
as the yet-unseen $c\bar c$ state $\chi_{c1}(2P)$, but its mass is
several tens of MeV lower than expected.  Moreover, $\Gamma_{X(3872)}
< 1.2$~MeV, while the $\chi_{c1}(1P)$, its ostensible radial ground
state, has a width $\Gamma = 0.84$~MeV, almost as large.  The
$\chi_{c1}(1P)$ has a mass $360$~MeV lower, so one expects the
$\chi_{c1}(2P)$ to have all of the $1P$ state's decay modes (as well
as many additional ones), but with much more phase space, and hence a
substantially larger width.

The $J^{PC} = 0^{++}$ state $\chi_{c0}^{\vphantom\dagger} (3915)$ is
an even trickier example.  Its mass lies very close to quark-potential
model predictions for that of the yet-unseen $c\bar c$ state
$\chi_{c0}^{\vphantom\dagger} (2P)$.  As described in detail below, it
is produced in $\gamma \gamma$ fusion, as one would expect for the
$\chi_{c0}^{\vphantom\dagger} (2P)$, and $\chi_{c0}^{\vphantom\dagger}
(3915)$ was briefly hailed by the Particle Data Group
(PDG)~\cite{Agashe:2014kda} as the missing $c\bar c$ state
$\chi_{c0}^{\vphantom\dagger} (2P)$.  However, the current absence of
the expected dominant $D^{(*)} \bar D^{(*)}$ decay modes speaks
against a $c\bar c$ interpretation, and indeed, also against a $c\bar
c q\bar q$ interpretation ($q = u, d$).

In this work, we therefore propose that $\chi_{c0}^{\vphantom\dagger}
(3915)$ is the lightest hidden-charm, hidden-strangeness ($c\bar c
s\bar s$) tetraquark state.  Our analysis is performed assuming the
diquark-antidiquark model first proposed in Ref.~\cite{Maiani:2004vq}
and applied to $c\bar c s\bar s$ states in Ref.~\cite{Drenska:2009cd}
(where the lightest $c\bar c s\bar s$ state was indeed found to have
$J^{PC} = 0^{++}$).  Since the advent of those two papers, many new
exotic states have been observed, and the model was improved recently
to reflect the new data in Ref.~\cite{Maiani:2014aja}.  Our analysis,
therefore, develops this improved version of the diquark model for
$c\bar c s\bar s$ states, under the assumption that
$\chi_{c0}^{\vphantom\dagger} (3915)$ is their ground state.

Along the way we predict the full spectrum of $c\bar c s\bar s$
states, noting several whose properties match those of observed
exotics remarkably well.  For example, the $Y(4140)$ observed in $B$
decays appears as an enhancement in the $J/\psi \, \phi$ spectrum,
exactly as expected for a $c\bar c s\bar s$ state, but it has not yet
appeared in $\gamma \gamma$ fusion experiments.  Our model neatly
accommodates a $J^{PC} = 1^{++}$ state at 4140~MeV, which is forbidden
by the Landau-Yang theorem~\cite{Landau:1948kw,Yang:1950rg} from
coupling to a two-photon state.

This paper is organized as follows.  In Sec.~\ref{sec:Chi3915} we
review the measured properties of the $\chi_{c0}^{\vphantom\dagger}
(3915)$ to motivate the proposal that it and several other exotics may
be $c\bar c s\bar s$ states.  Section~\ref{sec:Diquark} introduces the
diquark-antidiquark model used and develops its spectrum of $c\bar c
s\bar s$ states.  We analyze our results in Sec.~\ref{sec:Analysis} by
comparing to the known exotics spectrum, pointing out both successes
and shortcomings of the results.  In Sec.~\ref{sec:Concl} we present a
brief discussion and conclude.

\section{$\chi^{\protect\vphantom\dagger}_{c0}(3915)$ and Other
Potential $c\bar c s\bar s$ States} \label{sec:Chi3915}

An understanding of the exotic charmoniumlike spectrum remains
elusive, to say the least, from both experimental and theoretical
viewpoints (See Ref.~\cite{Chen:2016qju} for a thorough review and
Ref.~\cite{Briceno:2015rlt} for perspectives on future prospects.).
With respect to the current work, the most interesting state is
$\chi^{\vphantom\dagger}_{c0}(3915)$, which was discovered by Belle in
2005~\cite{Abe:2004zs} as a $J/\psi \, \omega$ enhancement in the
process $B \to J/\psi \, \omega K$ [and was originally labeled
$Y(3940)$], and confirmed by
BaBar~\cite{Aubert:2007vj,delAmoSanchez:2010jr}.  However, Belle found
no evidence for $D^{*0} \bar D^0$ decays of the
state~\cite{Adachi:2008sua}.  In 2010, Belle
discovered~\cite{Uehara:2009tx} the state $X(3915)$ in $\gamma \gamma
\to J/\psi \, \omega$, and BaBar subsequently confirmed the
result~\cite{Lees:2012xs}, establishing furthermore that the state has
$J^{PC} = 0^{++}$, so that its name under the conventional scheme
should be $\chi^{\vphantom\dagger}_{c0}$.  However, again, no evidence
for a peak near 3915~MeV in $D^{(*)} \bar D^{(*)}$ was found in $B \to
D^{(*)} \bar D^{(*)} K$ decays at Belle~\cite{Brodzicka:2007aa} or
BaBar~\cite{Aubert:2008gu}.  The shared $J/\psi \, \omega$ decay mode
and proximity in mass and width for these two states has led them to
be identified as the same state, currently called
$\chi^{\vphantom\dagger}_{c0}(3915)$.  Its mass and width are
currently given as~\cite{Agashe:2014kda}:
\begin{equation} \label{eq:3915facts}
M = 3918.4 \pm 1.2 \, {\rm MeV}, \hspace{1em} \Gamma = 20 \pm 5 \,
{\rm MeV} \, .
\end{equation}

In fact, the establishment of $J^{PC} = 0^{++}$ for
$\chi^{\vphantom\dagger}_{c0}(3915)$ immediately suggested that the
state is actually the first radial excitation
$\chi^{\vphantom\dagger}_{c0}(2P)$ of the known conventional
charmonium state $\chi^{\vphantom\dagger}_{c0}(1P)$, the $2P$ state
mass being predicted in quark potential models to lie in the range
3842--3916~MeV~\cite{Barnes:2005pb,Li:2009zu,Wang:2014voa}.  The $2P$
identification was also briefly espoused by the Particle Data Group
(PDG)~\cite{Agashe:2014kda} (in its online form).  However, this
identification was questioned by
Refs.~\cite{Guo:2012tv,Wang:2014voa,Olsen:2014maa}; their objections
amount to: i)~The mass splitting between the established
$\chi^{\vphantom\dagger}_{c2}(2P)$ (3927~MeV) and
$\chi^{\vphantom\dagger}_{c0}(3915)$ is rather smaller than expected
from quark potential models; ii)~The true $c\bar c$
$\chi^{\vphantom\dagger}_{c0}(2P)$ should decay copiously to $D^{(*)}
\bar D^{(*)}$ (the $D^0 \bar D^{*0}$ threshold lies at 3872~MeV, and
the $D^0 \bar D^0$ threshold lies at 3730~MeV); iii)~As a
charmonium-to-charmonium process, the decay
$\chi^{\vphantom\dagger}_{c0}(2P) \to J/\psi \, \omega$ is
Okubo-Zweig-Iizuka (OZI) suppressed and would be expected to occur
less frequently than is observed.  In fact, Ref.~\cite{Olsen:2014maa}
showed that the tension between ii) and iii) if
$\chi^{\vphantom\dagger}_{c0}(3915)$ is assumed to be
$\chi^{\vphantom\dagger}_{c0}(2P)$ leads to incompatible bounds on the
branching fraction ${\cal B}(\chi^{\vphantom\dagger}_{c0}(2P) \to
J/\psi \, \omega)$.  As a result of these objections, the PDG
currently refers to the state as $\chi^{\vphantom\dagger}_{c0}(3915)$.

Some comments regarding the $J^{PC}$ assignment in $\gamma \gamma$
fusion are in order.  If the photons are both transversely polarized,
then the Landau-Yang theorem~\cite{Landau:1948kw,Yang:1950rg} forbids
the resonance from having spin one.  Of course, the photons at Belle
and BaBar are produced from $e^+ e^-$ collisions, and
longitudinally-polarized off-shell photons can evade this constraint.
However, the photon virtuality in this case scales with $m_e$, which
is much smaller than the other mass scales in the process.  The
difference between the longitudinal and timelike photon polarizations
(the latter of which gives an exactly vanishing contribution to
physical amplitudes due to the Ward identity) then vanishes with
$m_e$, meaning that longitudinal photon contributions also vanish in
this limit.  Noting both $P$ and $C$ conservation in QED and using
Bose symmetry, the allowed quantum numbers for resonances formed in
$e^+ e^- \to \gamma \gamma \to X$ are therefore indeed either $0^{++}$
or $2^{++}$.

The $\chi^{\vphantom\dagger}_{c0}(3915)$ therefore appears to be a
supernumerary $0^{++}$ charmoniumlike state, and very likely a 4-quark
state (the lowest $0^{++}$ hybrid computed by lattice QCD being
expected to lie many hundreds of MeV higher~\cite{Braaten:2014qka}).
It is most natural to suppose that
$\chi^{\vphantom\dagger}_{c0}(3915)$ has the flavor structure of an
isosinglet: $c\bar c (u\bar u - d\bar d)/\sqrt{2}$.  Indeed, searches
for signals of charged partner states $c\bar c u\bar d$ or $c\bar c
d\bar u$~\cite{Aubert:2004zr,Choi:2011fc} in the same energy range
[actually designed to look for $X(3872)$ isospin partners] have
produced no clear signal.\footnote{We thank S.~Olsen for pointing out
  this very important fact.}  Furthermore, such a 4-quark state would
seem to have no obvious barrier for decaying into $D \bar
D$,\footnote{The only known exception to this statement is if the
  state is a molecule of two mesons held together primarily through
  $0^-$ exchanges, such as by $\pi$ and $\eta$.  In that case, Lorentz
  symmetry plus $P$ conservation of strong interactions forbids decay
  into two $0^-$ mesons.} and only have a relatively small $p$-wave
barrier for decay into $D \bar D^{*}$.  The absence of observed
open-charm decays of $\chi^{\vphantom\dagger}_{c0}(3915)$ poses a real
problem for the 4-quark interpretation.

We propose, therefore, a rather radical solution: The
$\chi^{\vphantom\dagger}_{c0}(3915)$ is a $c\bar c s\bar s$ state,
hence naturally an isosinglet that eschews open-charm decays.  It lies
just below the $D_s^+ \bar D_s^-$ threshold (3937~MeV) as well as the
$J/\psi \, \phi$ threshold (4116~MeV), and therefore the only
OZI-allowed decay (in that no new flavors in a quark-antiquark pair
are created or destroyed) open to it is $\eta_c \eta$ (threshold
3531~MeV).\footnote{Note that no exotic to $\eta_c$ decays have yet
  been observed~\cite{Vinokurova:2015txd}.}  We present a calculation
of this width in Sec.~\ref{sec:Analysis} and argue that it naturally
accommodates the value in Eq.~(\ref{eq:3915facts}).  The observed
decay mode $J/\psi \, \omega$ actually appears to be quite suppressed,
being either due to $\omega$-$\phi$ mixing that is less than ideal (so
that $\omega$ contains a small amount of valence $s\bar s$, and $\phi$
contains a small amount of valence $q\bar q$), or double
OZI-suppression ($s\bar s \to g \to q\bar q$).  Furthermore, we assert
that $\chi^{\vphantom\dagger}_{c0}(3915)$ is the lightest $c\bar c
s\bar s$ state; the only lighter charmoniumlike exotic is $X(3872)$,
and it decays freely into open-charm states.

A number of higher exotic states have properties amenable to a $c\bar
c s\bar s$ description, by virtue of having neither obvious isospin
partners nor observed open-charm decays.  Including the
$\chi^{\vphantom\dagger}_{c0}(3915)$, 9 states share these properties:
$Y(4008)$, $Y(4140)$, $Y(4230)$, $Y(4260)$, $Y(4274)$, $X(4350)$,
$Y(4360)$, and $Y(4660)$.  This list includes 4 of the 5 states,
$Y(4008)$, $Y(4260)$, $Y(4360)$, and $Y(4660)$, observed using
initial-state radiation (ISR) production in $e^+ e^-$ annihilation,
and therefore necessarily carrying $J^{PC} = 1^{--}$; the fifth,
$Y(4630)$, decays to $\Lambda_c^+ \bar \Lambda_c^-$.  $Y(4008)$ and
$Y(4260)$ have been seen only in decays containing a $J/\psi$, while
$Y(4360)$ and $Y(4660)$ have been seen only in decays containing a
$\psi(2S)$.  ISR states curiously also do not appear as obvious peaks
in the $R(e^+ e^- \to {\rm hadrons})$ ratio, unlike the conventional
$1^{--}$ charmonium states $J/\psi$, $\psi(2S)$, $\psi(3770)$,
$\psi(4040)$, $\psi(4160)$, and $\psi(4415)$~\cite{Agashe:2014kda}
(Indeed, a local minimum of $R$ appears around 4260~MeV).  If this
effect reflects the relative difficulty of making extra particles in
$e^+ e^-$ annihilation at energies where $\alpha_s$ is small [{\it
  i.e.}, with $\alpha_s(m_c) \simeq 0.3$, producing not just $c\bar
c$, but $c\bar c g$ or $c\bar c q\bar q$], then the production of
$c\bar c s\bar s$ would presumably be even further suppressed due to a
mass effect.

The $Y(4140)$, $Y(4274)$, and $X(4350)$ are even better $c\bar c s
\bar s$ candidates, since they are observed as $J/\psi \, \phi$
enhancements.  The $Y(4140)$ was first reported by CDF in the process
$B \to J/\psi \, \phi K$ in 2009~\cite{Aaltonen:2009tz}, and presented
with higher statistics by them in 2011~\cite{Aaltonen:2011at}, with
other observations in this channel provided by
D$\O$~\cite{Abazov:2013xda} and CMS~\cite{Chatrchyan:2013dma}, while
LHCb has not yet seen the state, but the disagreement is only at the
level of $2\sigma$~\cite{Aaij:2012pz}.  Along the way,
Refs.~\cite{Aaltonen:2011at,Chatrchyan:2013dma} observed in the same
channel the enhancement called $Y(4274)$.  Belle, however, using the
production mode $\gamma \gamma \to J/\psi \, \phi$, saw neither
$Y(4140)$ nor $Y(4274)$, but instead discovered a new state,
$X(4350)$~\cite{Shen:2009vs}.  A possible explanation for the absence
of $Y(4140)$ and $Y(4274)$ in $\gamma \gamma$ production is of course
the Landau-Yang theorem, granted that neither state is $J^{PC} =
0^{++}$ nor $2^{++}$.  A study of $Y(4140)$, $Y(4274)$, and $X(4350)$
as $c\bar c s\bar s$ states using QCD sum rules (but leading to rather
different $J^{PC}$ assignments) appears in Ref.~\cite{Wang:2015pea},
while Ref.~\cite{Stancu:2009ka} is a quark-model study predicting
$Y(4140)$ to be $1^{++}$ and notes the importance of the $\eta_c \eta$
mode.

Lastly, the $Y(4230)$ is an enhancement seen in the process $e^+ e^-
\to \chi_{c0} \, \omega$~\cite{Ablikim:2014qwy}. Should it turn out to
be a $c\bar c s \bar s$ state, its $\chi_{c0} \, \omega$ decay must
proceed through the same $\omega$-$\phi$ mixing or double-OZI
suppression mechanism as suggested for
$\chi^{\vphantom\dagger}_{c0}(3915)$.

\section{Diquark Models} \label{sec:Diquark}

Interest in diquark-antidiquark models for light scalar mesons has a
long and interesting history (see, {\it e.g.},
Ref.~\cite{Jaffe:2004ph} for a review).  The decay patterns for such
states obtained from the OZI rule are discussed in
Ref.~\cite{Maiani:2004uc}, and those from instanton-induced decays are
discussed in Ref.~\cite{Hooft:2008we}.  Here, however, we focus on an
approach obtained from simple Hamiltonian considerations.\footnote{
For example, studies of tetraquarks by allowing for flavor breaking
through chromomagnetic interactions have a long history in the
literature~\cite{Buccella:2006fn}.}

The ``Type-I'' diquark model of Ref.~\cite{Maiani:2004vq} is defined
in terms of a Hamiltonian with local spin-spin couplings combined with
spin-orbit and purely orbital terms.  The orbital angular momentum
operator ${\bf L}$ refers to the excitation between the
diquark-antidiquark pair, while orbital excitations within each
diquark are ignored.  Specializing (for notational simplicity) to
4-quark systems with hidden charm $[c q_1][\bar c \bar q_2]$, the
Hamiltonian reads
\begin{equation} \label{eq:H_full}
H = m_{[c q_1]} + m_{[\bar c \bar q_2]} + H_{SS}^{qq} + H_{SS}^{q
\bar q} + H_{SL} + H_L \, ,
\end{equation}
where $m_{[c q_1]}$ and $m_{[\bar c \bar q_2]}$ are the diquark
masses, $H_{SS}^{qq}$ refers to spin-spin couplings between two quarks
(or antiquarks) and therefore refers to spin-spin couplings within
either the diquark or antidiquark:
\begin{equation} \label{eq:HSSqq}
H_{SS}^{qq} = 2\kappa_{[c q_1]} \, {\bf s}_c \! \cdot {\bf s}_{q_1} +
2\kappa_{[\bar c \bar q_2]} \, {\bf s}_{\bar c} \! \cdot {\bf s}_{\bar
q_2} \, ,
\end{equation}
$H_{SS}^{q \bar q}$ couples quarks to antiquarks, and therefore
induces interactions between the diquark and the antidiquark:
\begin{eqnarray}
H_{SS}^{q \bar q} & = &
2 \kappa_{c \bar q_2}   \, {\bf s}_c     \! \cdot {\bf s}_{\bar q_2} +
2 \kappa_{c \bar c}     \, {\bf s}_c     \! \cdot {\bf s}_{\bar c}   
\nonumber \\ & & +
2 \kappa_{q_1 \bar c}   \, {\bf s}_{q_1} \! \cdot {\bf s}_{\bar c}   +
2 \kappa_{q_1 \bar q_2} \, {\bf s}_{q_1} \! \cdot {\bf s}_{\bar q_2}
\, , \label{eq:HSSqqbar}
\end{eqnarray}
and $H_{SL}$ and $H_L$ are the spin-orbit and purely orbital terms,
respectively:
\begin{eqnarray}
H_{SL} & = & -2a ({\bf s}_{[cq_1]} \! \cdot {\bf L} + {\bf s}_{[\bar c
  \bar q_2]} \! \cdot {\bf L}) = -2a \, {\bf S} \! \cdot
{\bf L} \, , \nonumber \\
H_L & = & \frac{B_c}{2} {\bf L}^2 \, , \label{eq:H_orb}
\end{eqnarray}
where ${\bf S}$ is the total quark spin operator.  The ``Type-II''
diquark model~\cite{Maiani:2014aja} is defined by neglecting all
spin-spin couplings between a quark of the diquark and an antiquark of
the antidiquark, {\it i.e.}, effectively by setting $H_{SS}^{q \bar q}
= 0$.  The dynamics binding tetraquark states can be very different
from that binding conventional hadrons, so one should not expect a
``universal'' Hamiltonian to hold for all hadrons.

The most natural basis in which to describe the diquark-antidiquark
states is one in which the good quantum numbers are the four quark
spins $s_c$, $s_{\bar s}$, $s_{q_1}$, $s_{\bar q_2}$, diquark spins
$s_{[cq_1]}$, $s_{[\bar c \bar q_2]}$, total quark spin $S$, orbital
angular momentum $L$, and total angular momentum $J$.  One can also
recouple the quark spins into $s_{c \bar c}$, $s_{q_1 \bar q_2}$ using
the Wigner $9j$ symbol~\cite{Edmonds:1996}.  With $q_1 = q_2 = s$,
\begin{eqnarray}
\lefteqn{\left< \left( s_s s_c \right) s_{[cs]} , \left( s_{\bar s}
s_{\bar c} \right) s_{[\bar c \bar s]}, JM \left| \right.
\left( s_s s_{\bar s} \right) s_{s \bar s} , \left( s_c s_{\bar c}
\right) s_{c \bar c} , JM \right>} & & \nonumber \\
& = & \sqrt{(2s_{[cs]}+1) (2s_{[\bar c \bar s]}+1) (2s_{s \bar s}+1)
(2s_{c \bar c}+1)} \hspace{2em} \nonumber \\
& & \times \left\{ \begin{array}{ccc}
s_s         & s_c         & s_{[cs]}            \\ 
s_{\bar s}  & s_{\bar c}  & s_{[\bar c \bar s]} \\
s_{s\bar s} & s_{c\bar c} & J
\end{array} \right\} \, . \label{eq:9j}
\end{eqnarray}
This basis is particularly convenient for identifying the charge
conjugation ($C$) quantum number of the states:
\begin{equation} \label{eq:Cparity}
C = (-1)^{s_{c \bar c} + s_{s \bar s} + L} \, .
\end{equation}

The $c\bar c s\bar s$ tetraquark states have received a dedicated
study only in the Type-I model~\cite{Drenska:2009cd}, some years ago
when many known exotic charmoniumlike states had not yet been
observed.  Should $c\bar c s\bar s$ tetraquark states be produced,
their natural OZI-allowed decays are the open-charm, open-strangeness
modes $D_s^{(*)} \bar D_s^{(*)}$ (if kinematically possible), or
hidden-charm, hidden-strangeness decays such as $J/\psi \, \phi$,
$\eta_c \eta$, {\it etc.}, depending upon the $J^{PC}$ of the state.
In particular, open-charm decays $D^{(*)} \bar D^{(*)}$ are expected
to be suppressed because they are doubly OZI suppressed: The $s\bar s$
pair must annihilate and a $q\bar q$ pair must be created.  As
discussed above, no less than 9 of the exotic charmoniumlike
candidates have not (yet) been seen to have open-charm decays:
$\chi_{c0}^{\vphantom\dagger} (3915)$, $Y(4008)$, $Y(4140)$,
$Y(4230)$, $Y(4260)$, $Y(4274)$, $X(4350)$, $Y(4360)$, and $Y(4660)$.
Furthermore, no exotic candidate has yet been seen to decay to
$D_s^{(*)} \bar D_s^{(*)}$.  The presence of possible $c\bar c s\bar
s$ states also ameliorates one of the more awkward problems of
tetraquark models: If hidden-charm tetraquarks contain light quarks,
then one expects either near-degenerate quartets $\{ c\bar c u\bar u$,
$c\bar c d\bar d$, $c\bar c u\bar d$, $c\bar c d\bar u \}$ or an
isosinglet-isotriplet combination of these states, all carrying the
same $J^{PC}$.  The original $X(3872)$ exotic discovered at
Belle~\cite{Choi:2003ue} is a $J^{PC} = 1^{++}$ state widely believed
to be $c\bar c q\bar q$, but dedicated searches for such partner
states~\cite{Aubert:2004zr,Choi:2011fc} have produced no
signal.\footnote{In the case of $X(3872)$, the absence of obvious
charged partners can be related to differing distances to the
isospin-partner neutral and slightly higher charged $D \bar D^*$
thresholds.  For example, the formation of the $X^\pm$ might be
suppressed by a Feshbach-type mechanism, as described
in~\cite{Esposito:2014rxa}.  Alternately, the natural level of the $X$
isotriplet states might be sufficiently high compared to the
largely-isosinglet $X(3872)$ that they may have escaped detection to
date due to having large widths.}  Of course, any states believed to
be $c\bar c s\bar s$ do not present this problem.

Implicit in these diquark models is the assumption of the validity of
a Hamiltonian approach, which in turn implies a single relevant time
coordinate (as the conjugate variable to the Hamiltonian), and hence a
common rest frame for the component quarks.  In reality, the quarks
can move relativistically, especially since the exotic states are
generally created in $b$-quark decays or colliders, in processes
accompanied by the release of large amounts of energy.  In particular,
the spin of a particle is measured in its rest frame, and therefore
the meaning of a spin-spin operator becomes obscured in highly
relativistic systems.  If needed, the mathematical way forward is to
employ a helicity formalism, as was most famously expounded in
Ref.~\cite{Jacob:1959at}.

From a dynamical point of view, one can imagine the heavy-quark
diquark and antidiquark to be fairly compact objects (tenths of a
fm)\footnote{In contrast, light-quark diquarks can be rather larger
[$O$(1)~fm]; for a lattice calculation, see~\cite{Alexandrou:2006cq}.}
that achieve a substantial separation (1~fm or more) due to the large
energy release, before being forced to hadronize due to confinement.
In this ``dynamical diquark picture''~\cite{Brodsky:2014xia}, the
implicit rest-frame approximations of
Refs.~\cite{Maiani:2004vq,Maiani:2014aja} are not wholly satisfactory,
but they should nevertheless provide a lowest-order set of
expectations for the spectrum of fully dynamical tetraquark states
produced via the diquark-antidiquark mechanism.  Moreover, the
dynamical diquark picture explains why exotics have only become
clearly visible in the heavy-quark sector: In the light-quark sector,
the diquark-antidiquark pair never achieve sufficient separation for
clear identification.  In the intermediate $s\bar s$ case, one may
discern some hints of diquark
structure~\cite{Lebed:2015fpa,Lebed:2015dca}.

Diquark structure, via the attractive channel of two color-{\bf 3}
quarks into a color-{$\bar{\bf 3}$} diquark, has also successfully
been used to explain the $P_c^+$ pentaquark states, both in the
original formulation~\cite{Maiani:2015vwa} and the dynamical
picture~\cite{Lebed:2015tna}.

With the formalism established, it is a simple matter to enumerate the
$[cs][\bar c \bar s]$ diquark-antidiquark states and compute their
masses using Eqs.~(\ref{eq:H_full})--(\ref{eq:H_orb}).  One finds the
6 $s$-wave and 14 $p$-wave states listed in Table~\ref{tab:states}.
The results in Table~I of Ref.~\cite{Drenska:2009cd} are analogous,
but once again, use a different model (as well as different numerical
inputs).  The mass formula obtained in the Type-II model is concise.
Since $q_1 = q_2 = s$, the diquark masses are equal, and only one
distinct spin-spin coupling $\kappa_{[cs]}$ appears:
\begin{eqnarray}
M & = & m_{[c q_1]} + m_{[\bar c \bar q_2]} + \frac{B_c}{2} L(L+1)
\nonumber \\ & & + a[L(L+1) + S(S+1) - J(J+1)] \nonumber \\
& & + \kappa_{[cs]} \left[ s_{[cs]} (s_{[cs]} + 1) +
s_{[\bar c \bar s]} (s_{[\bar c \bar s]} + 1) - 3 \right] \, .
\hspace{1em} \label{eq:Mass1}
\end{eqnarray}
Abbreviating
\begin{eqnarray}
M_0 & \equiv & m_{[c q_1]} + m_{[\bar c \bar q_2]} - 3\kappa_{[cs]} ,
\, \nonumber \\
\tilde B & \equiv & B_c + 2a , \, \nonumber \\
\alpha & \equiv & 2a , \, \nonumber \\
k & \equiv & 2\kappa_{[cs]} , \, \label{eq:newvars}
\end{eqnarray}
one immediately obtains
\begin{eqnarray}
M & = & M_0 + \frac{\tilde B}{2} L(L+1)
+ \frac{\alpha}{2} [S(S+1) - J(J+1)] \nonumber \\ & &
+ \frac{k}{2} \left[ s_{[cs]} (s_{[cs]} + 1) + s_{[\bar c \bar s]}
(s_{[\bar c \bar s]} + 1) \right] \, , \label{eq:Mass}
\end{eqnarray}
from which the mass expressions given in the last column of
Table~\ref{tab:states} follow.  The additional Type-I terms of
Eq.~(\ref{eq:HSSqq}) could also be computed, for example, by first
diagonalizing the states in a more convenient basis, using recoupling
formulas like Eq.~(\ref{eq:9j}); however, as seen in
Ref.~\cite{Drenska:2009cd}, the typical contributions from these terms
appear to be no more than about 20~MeV, which we can treat as a
systematic uncertainty in our mass predictions.  This uncertainty is
indicated henceforth by the use of the symbol ``$\simeq$''.

\begin{table}
  \caption{All $s$- and $p$-wave $c\bar c s\bar s$ diquark-antidiquark
    states.  In the cases $s_{[cs]} = 1$, $s_{[\bar c \bar s]} = 0$, linear
    combinations with $s_{[cs]} = 0$, $s_{[\bar c \bar s]} = 1$ states are
    understood to combine as needed [using Eqs.~(\ref{eq:9j}) and
    (\ref{eq:Cparity})] to form eigenstates of $C$.  State names used in
    Ref.~\cite{Maiani:2014aja} are also listed, and masses are obtained
    from Eq.~(\ref{eq:Mass}).\label{tab:states}}
\begin{tabular}{cl|cccc|l}
{\rm State} & \; $J^{PC}$ & $s_{[cs]}$ & $s_{[\bar c \bar s]}$ & $S$ &
\, $L$ \, & \hspace{2.5em} {\rm Mass} \\ \hline
$X_0$        & $\left| 0^{++} \right>_1$ & 0 & 0 & 0 & 0 &
$M_0$ \\
$X^\prime_0$ & $\left| 0^{++} \right>_2$ & 1 & 1 & 0 & 0 &
$M_0 + 2k$ \\
$X_1$        & $\left| 1^{++} \right>$   & 1 & 0 & 1 & 0 &
$M_0 + k$\\
$Z$          & $\left| 1^{+-} \right>_1$ & 1 & 0 & 1 & 0 &
$M_0 + k$ \\
$Z^\prime$   & $\left| 1^{+-} \right>_2$ & 1 & 1 & 1 & 0 &
$M_0 + 2k$ \\
$X_2$        & $\left| 2^{++} \right>$   & 1 & 1 & 2 & 0 &
$M_0 + 2k$ \\
             & $\left| 0^{--} \right>$   & 1 & 0 & 1 & 1 &
$M_0 + \tilde B + \alpha + k$ \\
             & $\left| 0^{-+} \right>_1$ & 1 & 0 & 1 & 1 &
$M_0 + \tilde B + \alpha + k$ \\
             & $\left| 0^{-+} \right>_2$ & 1 & 1 & 1 & 1 &
$M_0 + \tilde B + \alpha + 2k$ \\
$Y_1$        & $\left| 1^{--} \right>_2$ & 0 & 0 & 0 & 1 &
$M_0 + \tilde B -\alpha$ \\
$Y_2$        & $\left| 1^{--} \right>_1$ & 1 & 0 & 1 & 1 &
$M_0 + \tilde B +k$ \\
             & $\left| 1^{-+} \right>_1$ & 1 & 0 & 1 & 1 &
$M_0 + \tilde B +k$ \\
             & $\left| 1^{-+} \right>_2$ & 1 & 1 & 1 & 1 &
$M_0 + \tilde B+ 2k$ \\
$Y_3$        & $\left| 1^{--} \right>_3$ & 1 & 1 & 0 & 1 &
$M_0 + \tilde B -\alpha + 2k$ \\
$Y_4$        & $\left| 1^{--} \right>_4$ & 1 & 1 & 2 & 1 &
$M_0 + \tilde B + 2\alpha + 2k$ \\
             & $\left| 2^{-+} \right>_1$ & 1 & 0 & 1 & 1 &
$M_0 + \tilde B -2\alpha + k$ \\
             & $\left| 2^{-+} \right>_2$ & 1 & 1 & 1 & 1 &
$M_0 + \tilde B -2\alpha + 2k$ \\
             & $\left| 2^{--} \right>_1$ & 1 & 0 & 1 & 1 &
$M_0 + \tilde B -2\alpha + k$ \\
             & $\left| 2^{--} \right>_2$ & 1 & 1 & 2 & 1 &
$M_0 + \tilde B +2k$ \\
             & $\left| 3^{--} \right>$   & 1 & 1 & 2 & 1 &
$M_0 + \tilde B  -3\alpha + 2k$ \\
\end{tabular}
\end{table}

Using the results of Table~\ref{tab:states}, one can quickly establish
the mass hierarchy of states.  Assuming only that $k > 0$ [expected
from Eq.~(\ref{eq:HSSqq}) to hold, inasmuch as vector diquarks are
heavier than scalar diquarks], the lightest $s$-wave state is $X_0 =
\left| 0^{++} \right>_1$, which we naturally identify with the
$0^{++}$ state $\chi_{c0}^{\vphantom\dagger} (3915)$, and hence $M_0 =
3918.4$~MeV~\cite{Agashe:2014kda}$~\simeq 3920$~MeV\@.  One also
expects $\tilde B \ge 0$, or else orbitally excited states would
actually be lower in mass than $s$-wave states.  Lastly, the
spin-orbit coefficient $\alpha$ was argued in
Ref.~\cite{Maiani:2014aja} to be positive, so that masses increase
with $L$ and $S$ [as seen in Eq.~(\ref{eq:Mass1})]; an interesting
feature of this choice, as noted in Ref.~\cite{Cleven:2015era}, is
that with this inverted spin-orbit coupling, states of higher $J$ but
other quantum numbers the same are lighter [compare, {\it e.g.}, $Y_4
= \left| 1^{--} \right>_4$, $\left| 2^{--} \right>_2$, and $\left|
  3^{--} \right>$].

\section{Analysis} \label{sec:Analysis}

The strategy for the fit is now quite straightforward.  The $c\bar c
s\bar s$ spectrum depends upon 4 parameters: the multiplet base mass
$M_0$, the orbital excitation coefficient $\tilde B$, the spin-orbit
coefficient $\alpha$, and the diquark spin-spin coefficient $k$.  We
have noted that 9 candidate exotics may be used to fix these
parameters, and that the $s$- and $p$-wave bands consist of 20 states.
Therefore, the goal is to achieve a fit that predicts as many of the 9
exotics as possible, while not predicting any of the $20 - 9 =
11$~states with unseen $J^{PC}$ values to occur in mass regions where
they likely would already have been observed.

\subsection{Which $1^{--}$ States Are $c\bar c s\bar s$?}

Of particular note is that only 4 $1^{--}$ states occur in the $s$ and
$p$ waves; Ref.~\cite{Maiani:2014aja} notes that one more $1^{--}$
state, labeled $Y_5$, occurs in the $f$ wave ($s_{[cs]} = s_{[\bar c
  \bar s]} = 1$, $S=2$, $L=3$), but it is most likely much heavier
than the others considered here due to its high orbital excitation.
That being said, at least 4 $1^{--}$ candidate states have already
been observed in ISR processes: $Y(4008)$, $Y(4260)$, $Y(4360)$, and
$Y(4660)$ (although $Y(4008)$ has only been seen by
Belle~\cite{Yuan:2007sj,Liu:2013dau}).  In addition, the $Y(4230)$
seen by BESIII in $e^+ e^- \to \chi_{c0} \, \omega
$~\cite{Ablikim:2014qwy} is necessarily a $1^{--}$ state if formed in
the $s$ wave.  On the other hand, lattice calculations, while still
not in full agreement, concur that no more than one $1^{--}$
charmonium hybrid should occur below 4.5~GeV (see, {\it e.g.},
Ref.~\cite{Braaten:2014qka}, which predicts it to lie at a mass of
$4216 \pm 7$~MeV).

Also of note is that the neutral states so far
lacking open-charm decays appear to fall into at least two distinct
classes based upon their widths: Only $Y(4008)$ and $Y(4260)$ have
widths $> 100$~MeV\@.  One may suppose that one or both of these
states are $c\bar c q\bar q$ (hence possessing many more open channels
and thus a larger width) or $c\bar c g$ hybrids (so that OZI
suppression of $s\bar s$ annihilation is absent).  The $Y(4260)$ has
been observed in the 6-quark modes $J/\psi \, \pi \pi$ and $J/\psi \,
K^+ K^-$, which speaks against a hybrid structure, and the $J/\psi \,
\pi \pi$ channel speaks against a $c\bar c s\bar s$ structure [but see
further discussion on $Y(4008)$ later in this section].  In addition,
a recent study~\cite{Chen:2015dig} has calculated that the rate for
the radiative transition $Y(4260) \to \gamma X(3872)$ not only
suggests that $Y(4260)$ is a $c\bar c q\bar q$ state like $X(3872)$,
but also that both states are compatible with having the same
diquark-antidiquark wave function, except that $Y(4260)$ carries an
additional unit of orbital angular momentum.

\subsection{$s$-Wave States}

Of little ambiguity is the necessity of assigning
$\chi_{c0}^{\vphantom\dagger} (3915)$ the role of the $c\bar c s\bar
s$ band ground state $X_0 = \left| 0^{++} \right>_1$, which according
to Table~\ref{tab:states} immediately fixes the parameter $M_0 \simeq
3920$~MeV\@.  The full set of mass predictions is presented in
Table~\ref{tab:predict}.

\begin{table}
  \caption{Predictions for $s$-wave (the first 6 entries) and $p$-wave
  (the remaining 14 entries) $c\bar c s\bar s$ diquark-antidiquark
  state masses (in MeV), following the notation from
  Table~\ref{tab:states}.  Inputs are labeled by ``*''.  A ``?''
  indicates when more than one good assignment is possible.  In
  terms of Eq.~(\ref{eq:Mass}), the parameter fit values in MeV are
  $M_0 =3920$, $k = 220$, $\tilde B = \alpha =
  90$.}\label{tab:predict}
\begin{tabular}{cl|c|l}
{\rm State} & & {\rm Pred.\ Mass} & {\rm Observed} \\ \hline
$X_0$        & $\left| 0^{++} \right>_1$ & 3920 &
$\chi_{c0}^{\vphantom\dagger} (3915)^*$ \\
$X^\prime_0$ & $\left| 0^{++} \right>_2$ & 4360 & $X(4350)?$ \\
$X_1$        & $\left| 1^{++} \right>$   & 4140 & $Y(4140)^*$ \\
$Z$          & $\left| 1^{+-} \right>_1$ & 4140 & \\
$Z^\prime$   & $\left| 1^{+-} \right>_2$ & 4360 & \\
$X_2$        & $\left| 2^{++} \right>$   & 4360 & $X(4350)?$ \\
             & $\left| 0^{--} \right>$   & 4320 & $Y(4274)?$ \\
             & $\left| 0^{-+} \right>_1$ & 4320 & $Y(4274)?$ \\
             & $\left| 0^{-+} \right>_2$ & 4540 & \\
$Y_1$        & $\left| 1^{--} \right>_2$ & 3920 & $Y(4008)$ \\
$Y_2$        & $\left| 1^{--} \right>_1$ & 4230 & $Y(4230)^*$ \\
             & $\left| 1^{-+} \right>_1$ & 4230 & \\
             & $\left| 1^{-+} \right>_2$ & 4450 & \\
$Y_3$        & $\left| 1^{--} \right>_3$ & 4360 & $Y(4360)^*$ \\
$Y_4$        & $\left| 1^{--} \right>_4$ & 4630 & $Y(4660)$ \\
             & $\left| 2^{-+} \right>_1$ & 4050 & \\
             & $\left| 2^{-+} \right>_2$ & 4270 & $Y(4274)?$ \\
             & $\left| 2^{--} \right>_1$ & 4050 & \\
             & $\left| 2^{--} \right>_2$ & 4450 & \\
             & $\left| 3^{--} \right>$   & 4180 & \\
\end{tabular}
\end{table}

Beyond this start, however, hints from the exotic state decay modes
become essential.  Perhaps the other states most essential to describe
as $c\bar c s \bar s$ are those observed to decay into $J/\psi \,
\phi$, namely, $Y(4140)$, $Y(4274)$, and $X(4350)$.  Assuming that
$Y(4140)$ is the $X_1 = \left| 1^{++} \right>$, then using
Table~\ref{tab:states} one chooses $k \simeq 220$~MeV, which not only
resolves the absence of this state from $\gamma \gamma$ production via
the Landau-Yang theorem, but also allows immediate prediction of all
the other masses in the $s$-wave band.  In particular, one finds a
degenerate state $Z = \left| 1^{+-} \right>_1$ at 4140~MeV and another
$Z^\prime = \left| 1^{+-} \right>_2$ at 4360~MeV; note that the known
neutral isotriplet $J^{PC} = 1^{+-}$ states $Z_c^0(4025)$,
$Z_c^0(3900)$ lie rather lower in mass.  Additionally, one finds two
more degenerate states at 4360~MeV, $X_0^\prime = \left| 0^{++}
\right>_2$ and $X_2 = \left| 2^{++} \right>$.  Either of these is an
excellent candidate for the $X(4350)$ found in $\gamma \gamma$
production.

Returning to the $Y(4140)$, one may use Eq.~(\ref{eq:9j}) to find that
the state $X_1$ has solely $s_{c\bar c} = s_{s\bar s} = 1$ content.
At the quark level, one expects $\gamma \gamma$ fusion to produce one
of the quark-antiquark pairs first (and necessarily with $J^{PC} =
0^{++}$ or $2^{++}$), and the other pair to be produced as the result
of bremsstrahlung from one of the initial quarks.  Thus, even at the
quark level, one sees the production of such a state to be
problematic.

According to Table~\ref{tab:states} or \ref{tab:predict}, the $s$-wave
states are highly degenerate and obey a simple equal-spacing rule (in
$k$).  Note that no $s$-wave state therefore carries a mass close to
that of $Y(4274)$, reported by CDF~\cite{Aaltonen:2011at} as
$4274.4^{+8.4}_{-6.7} \pm 1.9$~MeV, and by
CMS~\cite{Chatrchyan:2013dma} as $4313.8 \pm 5.3 \pm 7.3$~MeV\@.
Fitting to the $p$-wave states requires input from the ISR state
masses, as discussed below.  Then, the sole potential candidate for
the first mass is $\left| 2^{-+} \right>_2$ at 4270~MeV, while the
second mass can be accommodated by either $\left| 0^{--} \right>$ or
$\left| 0^{-+} \right>_1$ at 4320~MeV\@.  In the first case, a lighter
$\left| 2^{-+} \right>_1$ state occurs at 4050~MeV, which lies below
the 4116~MeV $J/\psi \, \phi$ threshold and therefore could easily
have escaped detection to now.  In all cases, however, the fact that
none of these states have $J^{PC} = 0^{++}$ or $2^{++}$ means that
they cannot be created in $\gamma \gamma$ fusion, in agreement with
observation.

Before leaving the $s$-wave band, let us note interesting properties
of the $\chi_{c0}^{\vphantom\dagger} (3915)$ under this assignment.
We have seen in the previous section that its mass lies just below the
$D_s^+ \bar D_s^-$ threshold 3937~MeV\@.  However, it is extremely
problematic to identify $\chi_{c0}^{\vphantom\dagger} (3915)$ as a
$D_s \bar D_s$ molecule (which was proposed in Ref.~\cite{Li:2015iga})
held together by meson exchanges, again using a fact noted in the
previous section: $D_s^+$ and $\bar D_s^-$ are $J^P = 0^-$ states, and
coupling to a $0^-$ meson (presumably $\eta$) is forbidden by Lorentz
symmetry plus $P$ invariance.  Should $\chi_{c0}^{\vphantom\dagger}
(3915)$ prove to be a $c\bar c s\bar s$ state, it is almost certainly
not a hadronic molecule.  The closeness of the
$\chi_{c0}^{\vphantom\dagger} (3915)$ mass to the $D_s^+ \bar D_s^-$
threshold need not be considered an unnatural coincidence, as the
so-called ``cusps'' due to such thresholds have been shown to be
effective in attracting nearby states, in particular for heavy-quark
states~\cite{Bugg:2008wu,Blitz:2015nra}.

Second, we have noted that the only OZI-allowed and phase-space
allowed decay mode for a $c\bar c s\bar s$ state of this mass is
$\eta_c \eta$.  We propose that this is the dominant
$\chi_{c0}^{\vphantom\dagger} (3915)$ decay mode.  The recombination
of quark spins for the $X_0$ state according to Eq.~(\ref{eq:9j})
gives
\begin{equation} \label{eq:X0}
X_0 = \frac 1 2 \left| s_{c\bar c} = 0, s_{s\bar s} = 0 \right> +
\frac{\sqrt{3}}{2} \left| s_{c\bar c} = 1, s_{s\bar s} = 1 \right>
\, ,
\end{equation}
meaning that the $J/\psi$ modes, if kinematically allowed, are more
probable by a factor 3.  Likewise, the $\eta$ wave function is only
fractionally $s\bar s$:
\begin{equation} \label{eq:eta}
\eta = \frac{1}{\sqrt{6}} \left( \left| u\bar u \right>
+ \left| d\bar d \right> - 2 \left| s\bar s \right> \right) \, .
\end{equation}
The decay $\chi_{c0}^{\vphantom\dagger} (3915) \to \eta_c \eta$ is
otherwise a simple 2-body decay of a scalar to two (pseudo)scalars, and
therefore its width is of the form
\begin{equation}
\Gamma = \left| {\cal M} \right|^2 \frac{p}{8\pi M^2} \, ,
\end{equation}
where $M$ is the $\chi_{c0}^{\vphantom\dagger} (3915)$ mass and $p =
665.0$~MeV is the magnitude of the spatial momentum for the 2-body
decay.  The invariant amplitude ${\cal M}$ is seen to have dimensions
of mass; with $\Gamma = 20$~MeV, one finds $|{\cal M}| = 3.4$~GeV\@.
When the suppression factors suggested by
Eqs.~(\ref{eq:X0})--(\ref{eq:eta}) are removed, the ``natural''
amplitude for the process is about 8.3~GeV, a substantial number that
suggests the sole decay already observed,
$\chi_{c0}^{\vphantom\dagger} (3915) \to J/\psi \, \omega$, can occur
at a reasonable rate if the $\omega$ contains a phenomenologically
acceptable $s\bar s$ component.  For example, if the non-ideal mixing
$\epsilon$ of $\omega$ is parametrized as
\begin{equation} \label{eq:nonideal}
\omega = \cos \epsilon \frac{1}{\sqrt 2} \left( \left| u \bar u
\right> + \left| d \bar d \right> \right) + \sin \epsilon \left|
s \bar s \right> \, ,
\end{equation}
then using Eq.~(\ref{eq:X0}) and the same value of $|{\cal M}|$,
one finds $\Gamma (\chi_{c0}^{\vphantom\dagger} (3915) \to J/\psi \,
\omega) = 29.9 \sin^2 \! \epsilon$~MeV, which for, {\it e.g.},
$\epsilon = 10^{-3}$ gives $\Gamma = 29.9$~eV.

As mentioned above, the size of the $J/\psi \, \omega$ branching
fraction for $\chi_{c0}^{\vphantom\dagger} (3915)$, given in
Ref.~\cite{Uehara:2009tx} in the form
\begin{eqnarray}
& & \Gamma(\chi_{c0}^{\vphantom\dagger} (3915) \to \gamma \gamma)
\times {\cal B} (\chi_{c0}^{\vphantom\dagger} (3915) \to J/\psi \,
\omega ) \nonumber \\
& = & (61 \pm 17 \pm 8) \; {\rm eV} \, ,
\end{eqnarray}
is considered too large to be compatible with the expected size of
OZI-suppressed decays of conventional charmonium.  If
$\chi_{c0}^{\vphantom\dagger} (3915)$ is a $c\bar c s\bar s$ state,
then OZI violation is evaded if the decay mode is accomplished through
the presence of a small valence $s\bar s$ component in the $\omega$,
which means non-ideal $\omega$-$\phi$ mixing.  This effect has been
considered in heavy-quark decays such as $D_s^+ \to \omega e^+
\nu_e$~\cite{Gronau:2009mp}.  It might, however, be more complicated
in the 4-quark environment in the sense that $\omega$-$\phi$ mixing
influenced by final-state interactions can have a significantly
different strength than in exclusive processes in which $\omega$ is
the only hadron present.

\subsection{$p$-Wave States}

Let us now turn to the $p$ waves.  We have already fixed 2 of the 4
model parameters, $M_0$ and $k$, from the $s$ waves.  When including
the $p$ waves, we find that the fits best representing the known
spectrum and introducing fewer light unknown states leave out
$Y(4260)$ and keep $Y(4008)$.  We have remarked above that these are
the two widest neutral charmoniumlike states, and are therefore the
best candidates for $c\bar c q\bar q$, and also that the mode
$Y(4260)$ in particular is very unlikely to be purely $c\bar c s\bar
s$.  Therefore, in the fit we present in Table~\ref{tab:predict}, the
$Y(4260)$ is excluded.

It should however be noted that the $Y(4008)$, which has only been
seen by Belle~\cite{Yuan:2007sj,Liu:2013dau} is even wider ($M =
3890.8 \pm 40.5 \pm 11.5$~MeV, $\Gamma = 254.5 \pm 39.5 \pm 13.6$~MeV,
according to Ref.~\cite{Liu:2013dau}), and like $Y(4260)$, decays to
$J/\psi \, \pi \pi$ (indeed, they are seen together in the same
experiment).  However, note that the central value for the $Y(4008)$
mass actually lies lower than that of the
$\chi_{c0}^{\vphantom\dagger} (3915)$ and well above the thresholds
for the $p$-wave $c\bar c s\bar s$ modes $\eta_c \eta$ (again,
3531~MeV) and $J/\psi \, \eta$ (3645~MeV), as well as the
$\omega$-$\phi$ mixing modes, $\eta_c \, \omega$ (3766~MeV) and
$J/\psi \, \omega$ (3880~MeV).  However, $Y(4008)$ lies well below the
$J/\psi \, \phi$ threshold (4116~MeV) but $Y(4260)$ lies well above
it; if $Y(4260)$ contained a substantial $c\bar c s\bar s$ component,
presumably its $J/\psi \, \phi$ mode would have been prominently
observed.

The closeness of the $Y(4008)$ and $\chi_{c0}^{\vphantom\dagger}
(3915)$ masses has an additional peculiar effect.  If one identifies
$Y(4008)$ as the lightest $J^{PC} = 1^{--}$ $c\bar c s \bar s$ state
$Y_1 = \left| 1^{--} \right>_2$, then the fit in
Table~\ref{tab:predict} gives $\tilde B = \alpha$, or using
Eq.~(\ref{eq:newvars}), $B_c = 0$ in the original notation of
Eq.~(\ref{eq:H_orb}), which means that the only orbital coupling
appears through the spin-orbit term.

In fact, the actual fit in Table~\ref{tab:predict} does not choose
$Y(4008)$ as an input, but rather chooses $Y(4230) = Y_2 = \left|
  1^{--} \right>_1$ and $Y(4360) = Y_3 = \left| 1^{--} \right>_3$ to
fix $\tilde B = \alpha = 90$~MeV\@.  Then, the prediction of $Y(4008)$
as $Y_1$ and $Y(4660)$ as $Y_4 = \left| 1^{--} \right>_4$ is
noteworthy.  An additional feature commending this choice is that
Eq.~(\ref{eq:9j}) can again be used to show that $Y_2$ contains only
terms in which $s_{c\bar c} = s_{s\bar s} = 1$, very much in agreement
with the $Y(4230)$ so far being seen only in the $\chi_{c0} \, \omega$
channel: The preferred decay mode would be $\chi_{c0} \, \phi$, but
its threshold is 4434~MeV, so again we suggest that $Y(4230)$ is a
$c\bar c s\bar s$ state that can decay via $\omega$-$\phi$ mixing.

Since the $D_s^{(*)+} \bar D_s^{(*)-}$ thresholds occur at 3937~MeV,
4081~MeV, and 4224~MeV, one would expect these ``fall-apart'' modes to
be the dominant ones for many of these states, particularly higher
ones such as $Y(4660)$.  However, it is worth noting that the best
current data for $e^+ e^- \to D_s^{(*)+} \bar
D_s^{(*)-}$~\cite{Pakhlova:2010ek} is only sensitive to the
conventional charmonium $\psi$ states; none of the exotics have yet
been seen to decay to charm-strange states.  Moreover, should the
dynamical diquark picture~\cite{Brodsky:2014xia} hold, such that more
highly energetic states entail greater separation of the diquarks and
therefore suppressed hadronization matrix elements, one then has a
natural mechanism for suppressing their decay widths beyond naive
expectations.

Lastly, this work presents only one of many possible fits to the known
exotic states lacking open-charm decays.  Several other possibilities
can occur, such as, {\it e.g.}, identifying the high-mass $1^{--}$
$Y(4660)$ state as the first in the $f$-wave ($L=3$) band ($s_{[cs]} =
s_{[\bar c \bar s]} = 1$, $S=2$, called $Y_5$ in
Ref.~\cite{Maiani:2014aja}).

\section{Conclusions} \label{sec:Concl}

Based on interesting patterns in the phenomenology of the
charmoniumlike states observed to date, we propose that the $J^{PC} =
0^{++}$ state $\chi_{c0}^{\vphantom\dagger} (3915)$ is the lightest
$c\bar c s\bar s$ state.  Its lack of observed $D^{(*)}\bar D^{(*)}$
decays argue against it being either the conventional $c\bar c$ state
$\chi^{\vphantom\dagger}_{c0}(2P)$ or a light-quark containing $c\bar
c q\bar q$ exotic state, and its single known decay mode $J/\psi \,
\omega$ can be understood as the $\omega$ having a small
(non-ideal mixing) $s\bar s$ component.

Furthermore, as a $c\bar c s\bar s$ state lying slightly below the
$D_s \bar D_s$ threshold, the $\chi_{c0}^{\vphantom\dagger} (3915)$ is
very unlikely to be a loosely bound molecule, and we therefore analyze
it as a diquark-antidiquark state.  Indeed, a state with $J^{PC} =
0^{++}$ in the mass region $\sim 3900$~MeV is precisely where the
lightest $c\bar c s\bar s$ state was expected in previous studies.
Importantly, even if $\chi_{c0}^{\vphantom\dagger} (3915)$ turns out
not to be $c\bar c s\bar s$, states with this quark content should
certainly appear in the same mass range as some of those already
observed.  To emphasize: One expects $c\bar c s\bar s$ states to occur
in the same range as other charmoniumlike states; and even if the
particular assignments in this paper are later disfavored, the
analysis leading to Table~\ref{tab:states} still holds.

Under the current hypothesis, however, some remarkable identifications
arise.  The $Y(4140)$, a $J/\psi \, \phi$ enhancement seen in $B$
decays, is naturally a $1^{++}$ $c\bar c s\bar s$ state which, by the
Landau-Yang theorem, is naturally absent from $\gamma \gamma$
production experiments (as is the case).  The $X(4350)$, $Y(4274)$,
and several of the $J^{PC} = 1^{--}$ $Y$ states arise naturally at
masses predicted for $c\bar c s\bar s$ states, and no unwanted extra
states that would already likely have been observed appear to occur.

The most flexible part of the identification---both experimentally and
theoretically---occurs in the $1^{--}$ sector: If so many of these
states are $c\bar c s\bar s$, what has happened to the expected $c\bar
c q\bar q$ states?  We have argued that $Y(4260)$ is almost certainly
$c\bar c q\bar q$ and is quite broad; one can imagine that the higher
ones are broader still, and thus difficult to discern. Indeed, the
very broad $Y(4008)$ might also be $c\bar c q\bar q$, and either the
true lowest $1^{--}$ $c\bar c s\bar s$ state is obscured by it, or
does not occur until it appears as $Y(4230)$. In any case, subsequent
experiments will certainly clarify the true nature of the full
spectrum, and $c\bar c s\bar s$ states will certainly play a role.

During the finalization of this paper, D$\O$
announced~\cite{D0:2016mwd} the observation of a new state in the
channel $B_s^0 \pi^\pm$, while a preliminary analysis by LHCb found no
evidence for such a state~\cite{LHCb:2016}.  Such a novel exotic
flavor structure, a tetraquark with only one heavy quark ($b\bar s
u\bar d$ for $\pi^+$), is expected to produce two states close in mass
(with $J^P = 0^+$, $1^+$) due to heavy-quark fine structure.  In
particular, if confirmed, it would be the first tetraquark not simply
of the $b\bar b q\bar q$ or $c\bar c q\bar q$ type, which makes
studies of new flavor structures like $c\bar c s\bar s$ all the more
timely.  Indeed, Ref.~\cite{D0:2016mwd} suggests the same type of
tetraquark paradigm as discussed here as being the most likely
structure.

\begin{acknowledgments}
  This work was supported by the National Science Foundation under
  Grant No.\ PHY-1403891 (RFL).  In addition, RFL thanks S.~Olsen for
  important comments.
\end{acknowledgments}


\end{document}